\documentclass[english,nolineno,a4paper]{socg-lipics-v2021}
\hideLIPIcs 

\usepackage[numbers,sort&compress]{natbib}

\title{Reconfiguration of 3D Pivoting Modular Robots}
\author{Hugo A. Akitaya}{University of Massachusetts Lowell,  USA}{}{}{}
\author{Frederick Stock}{University of Massachusetts Lowell,  USA}{}{}{}

\ccsdesc[500]{Theory of computation~Computational geometry}

\keywords{modular reconfigurable robotics, pivoting model, reconfiguration}

\Copyright{Hugo A. Akitaya and Frederick Stock}

\authorrunning{H.~A.~Akitaya and F.~Stock} 

\begin{document}

\maketitle

\begin{abstract}
\label{abstract}
We study a new model of 3-dimensional modular self-reconfigurable robots Rhombic Dodecahedral (RD).
By extending results on the 2D analog of this model we characterize the free space requirements for a pivoting move and investigate the \emph{reconfiguration problem}, that is, given two configurations $s$ and $t$ is there a sequence of moves that transforms $s$ into $t$? We show reconfiguration is PSPACE-hard for RD modules in a restricted pivoting model.
In a more general model, we show that RD configurations are not universally reconfigurable despite the fact that their 2D analog is [Akitaya et al., SoCG 2021]. Additionally, we present a new class of RD configurations that we call \textit{super-rigid}. Such a configuration remains rigid even as a subset of any larger configuration, which does not exist in the 2D setting.
\end{abstract}

\section{Introduction}
\label{Intro}

\emph{Programmable matter} refers to matter with the ability to change
its physical properties, such as shape, on demand. 
A popular approach to implement such a system is through \emph{modular self-reconfigurable robotic systems} (MSR) where small robotic units called \textit{modules} can attach and detach from each other, communicate, and move relative to each other, changing the shape of the system.
We require configurations to be \emph{connected}, meaning the adjacency graph of modules is connected (known as the \emph{single backbone condition}~\cite{dumitrescu2004motion}). 
A \emph{move} is an operation that transforms a configuration into another by changing the position of a single module. 
We focus on the \emph{pivoting model} where the moving module rotates around a static module.
This model is similar to many hardware implementations~\cite{romanishin2013m,piranda2018designing} but it is challenging from an algorithmic perspective as a single pivoting move can collide with several modules.

\begin{figure}[ht]
    \centering
    \includegraphics[width=.8\textwidth]{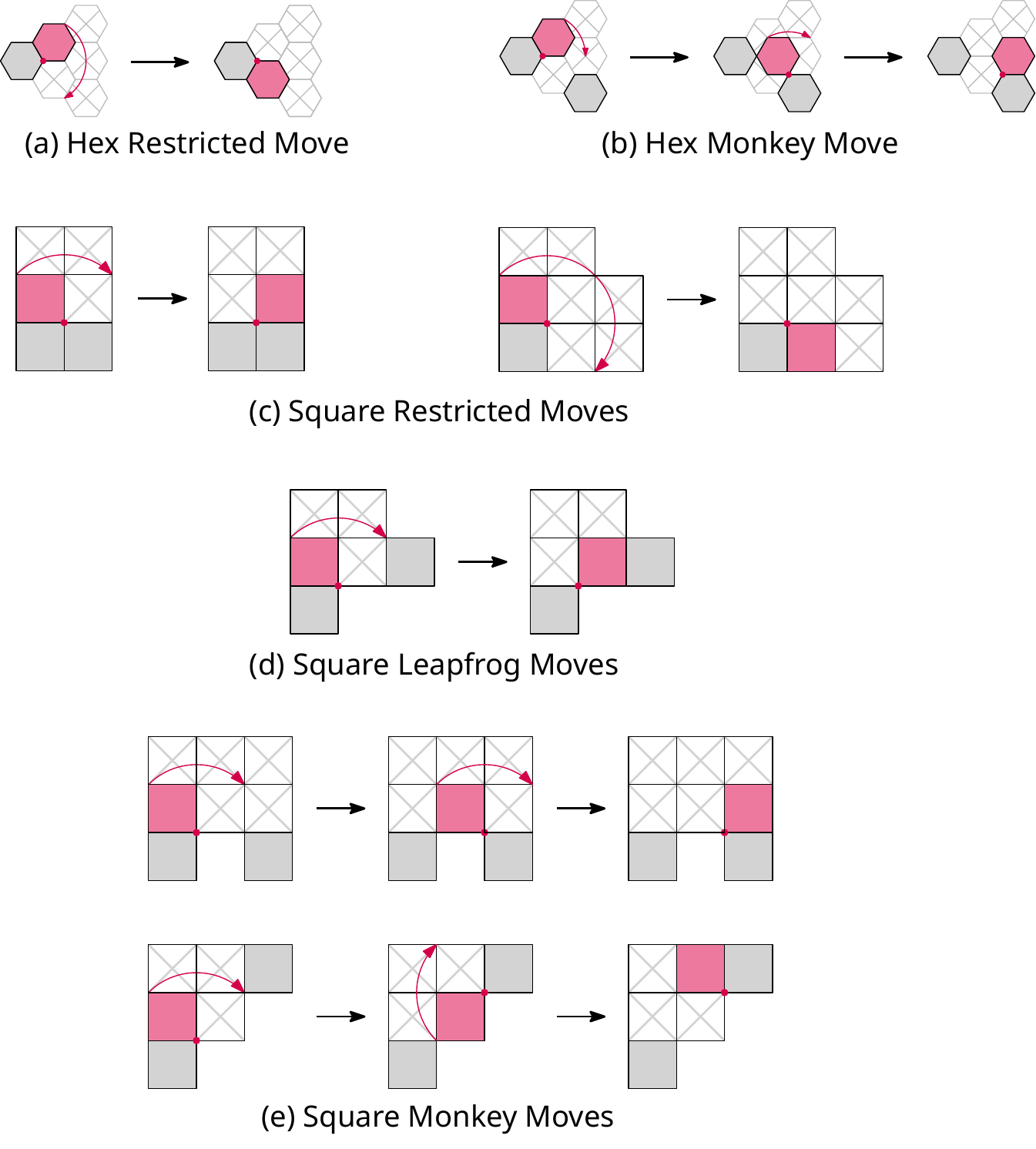}
    \caption{2-dimensional (a,c) restricted moves, (d) leapfrog, and (b, e) monkey moves.}
    \label{fig:plane-moves}
\end{figure}

If certain local configurations are forbidden, there are polynomial-time algorithms for square, hexagonal, and cube modules that compute a sequence of pivoting moves to transform a configuration into another~\cite{DBLP:conf/icra/SungBRR15, DBLP:journals/ral/FeshbachS21}. 
Akitaya et al.~\cite{Reconfiguration,Musketeers} classified three sets of pivoting moves for square modules and two for hexagonal modules and described their required free space, Fig.~\ref{fig:plane-moves}.
The \emph{restricted hexagonal model} only uses the restricted move while the \emph{monkey hexagonal model} uses both restricted and monkey moves.
The general problem was shown to be PSPACE-hard for the restricted hexagonal model~\cite{Reconfiguration}.
However, \cite{Reconfiguration} also gave a universal reconfiguration algorithm for the monkey hexagonal model that produces move sequences of length $O(n^3)$, for configurations of $n$ robots.

Not much is known about general reconfiguration in 3D models. 
A candidate to generalize hexagonal modules is the rhombic dodecahedron (RD). 
Implementations of RD-like modules exist~\cite{piranda2018designing}, but free space constraints for pivoting moves of these modules have never been described as in~\cite{Reconfiguration,Musketeers}.
We present such free space constraints to facilitate the description of 3D configurations and moves of such models.
We then generalize the PSAPCE-hardness proofs from~\cite{Reconfiguration} to RD pivoting models.
Finally, we show there are configurations of RD that are rigid even if they are a subset of a larger configuration, implying the configuration space of RD is disconnected. 

\section{Free-Space}
\label{FreeSpace}

\begin{figure}[h!]
    \centering
    \includegraphics[scale=.9]{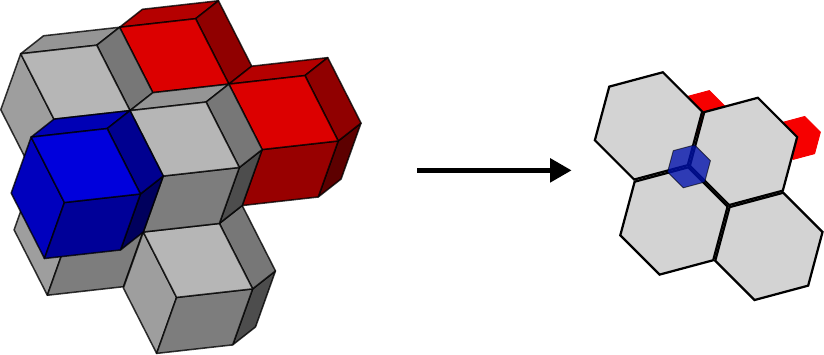}
    \caption{Equivalency between rhombic dodecahedra and hexagons. Red and blue modules are drawn small for legibility but are, in reality as large as the central grey modules. }
    \label{fig:RD2HEX}
\end{figure}

\begin{figure}[h!]
    \centering
    \includegraphics[width=.9\textwidth]{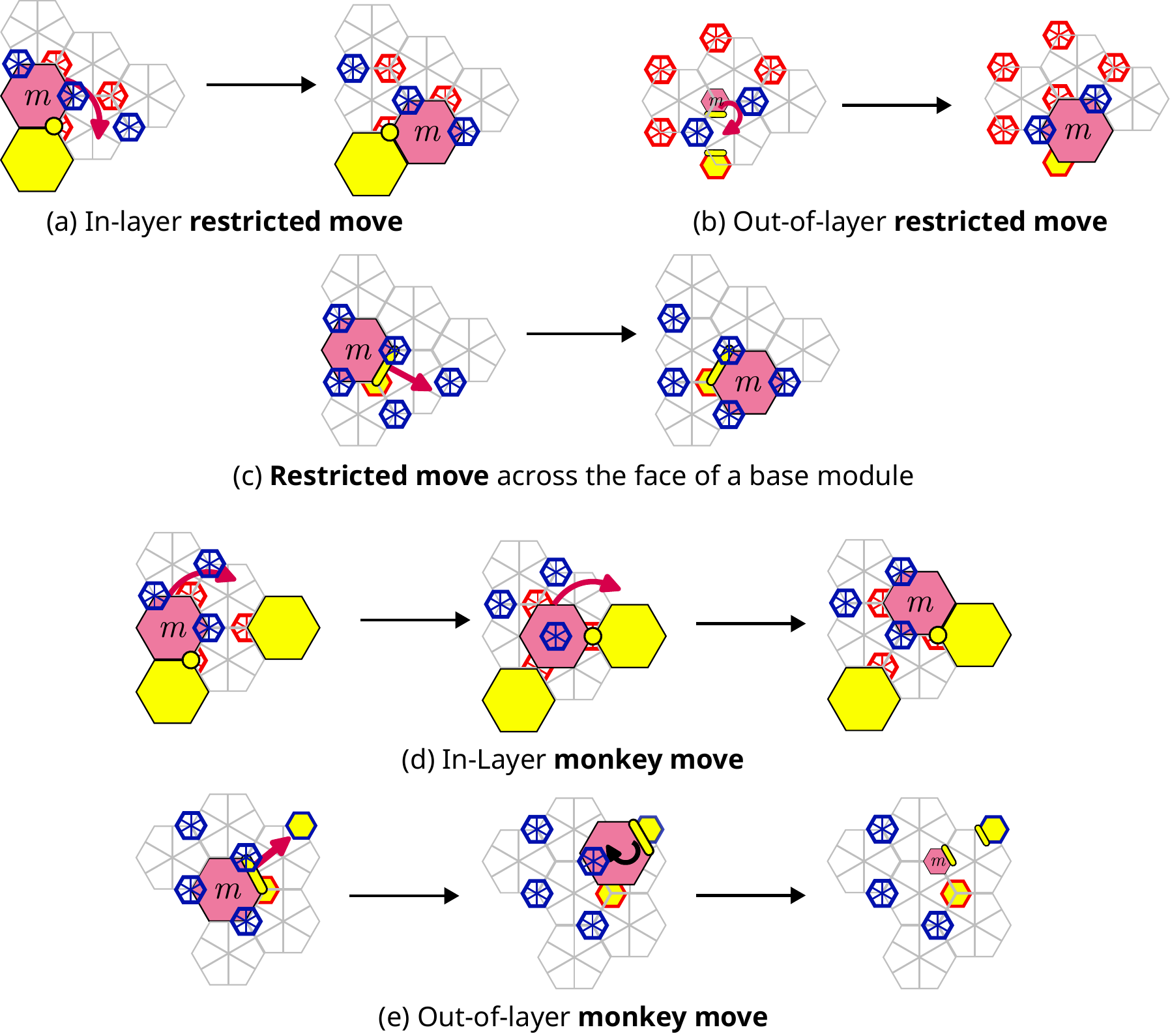}
    \caption{3-dimensional RD Moves. The moving module $m$ is pink, and the modules and the edges that $m$ pivots around are highlighted in yellow. The required free lattice positions (resp. top, resp. bottom) are marked with a grey (blue, red) asterisk.}
    \label{fig:RDMoves}
\end{figure}

%
A RD lattice can be represented as stacked layers of hexagons (Fig.~\ref{fig:RD2HEX}), and 
RD are space-filling polyhedra and therefore are a logical 3D analog to hexagonal MSR. 
We define restricted and monkey moves as in~\cite{Reconfiguration}. During a pivoting move, a RD module might collide with modules that are in layers adjacent to its plane of movement, so three layers of modules are needed to describe their free-space requirements, presented in Fig.~\ref{fig:RDMoves}.
A RD lattice is 3-cyclic so the bottom and top layer free-space requirements may flip based on a module's lattice position or direction of movement, hence Fig.~\ref{fig:RDMoves} is presented w.l.o.g.

\section{Reconfiguration is Hard}
\label{Hardness}

\begin{figure}[ht]
    \centering
    \includegraphics[width = .8\textwidth]{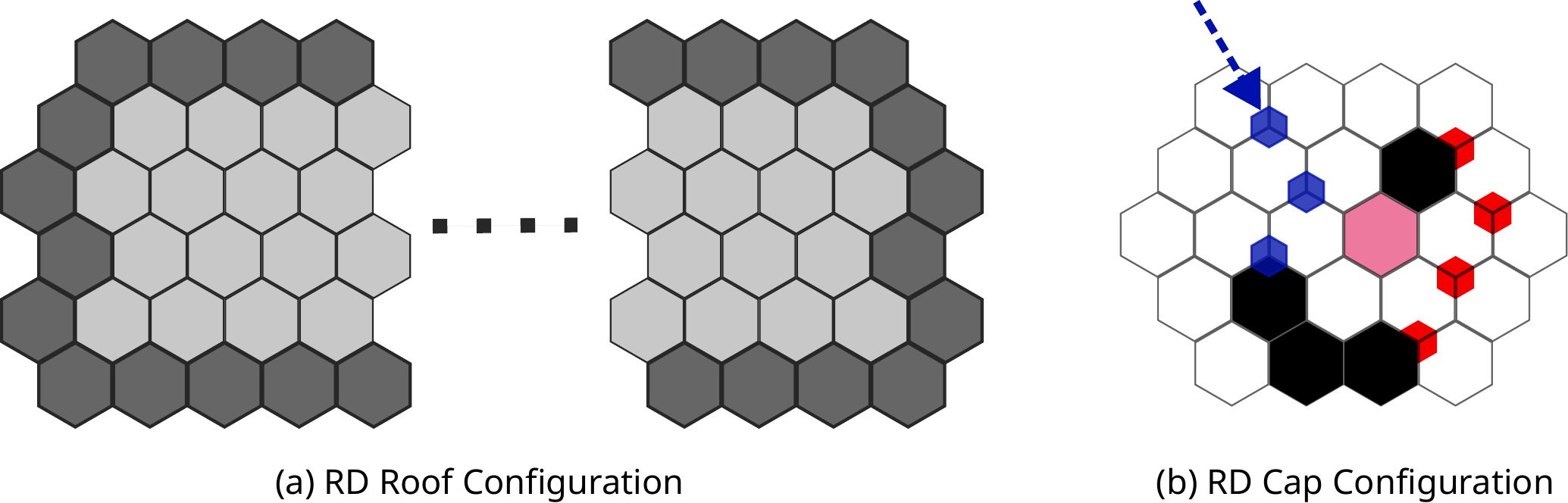}
    \caption{Roof and Cap configurations for rhombic dodecahedron.}
    \label{fig:roofs}
\end{figure}

\begin{theorem}
General Reconfiguration of RD with Restricted moves is PSPACE-hard.
\end{theorem}
We use the configurations in Fig.~\ref{fig:roofs} to extend the gadgets used by Akitaya et al. \cite{Reconfiguration} from 2D to 3D. 
A roof configuration is a gadget contained in a single layer formed by a cycle of modules, the \emph{boundary}, and all positions that are enclosed by the boundary. 
A cap configuration takes a path of modules and makes one end rigid. We use the cap to make our roof rigid by attaching a path of modules to each module in the boundary of a roof and terminating it with a cap.
By placing instances of the roof configurations two layers above and below the gadgets, no module can move up or down a layer, essentially restricting the gadgets to two dimensions. 
A similar approach can extend the square results from \cite{Reconfiguration} to prove reconfiguration of cubes under restricted, leapfrog, and monkey moves is PSPACE-hard as well. Details will be in an upcoming full version. 



\section{Rhombic Dodechahedra Are Super Rigid}
\label{SuperRigid}
In this section, we define a new class of configuration we call \emph{super-rigid}. 
\begin{definition}
A configuration $G$ is super rigid if given any configuration $C$ where $G$ is a subconfiguration, the modules in $G$ cannot be moved.  
\end{definition} 

\begin{figure}[ht]
    \centering
    \includegraphics[scale=0.75]{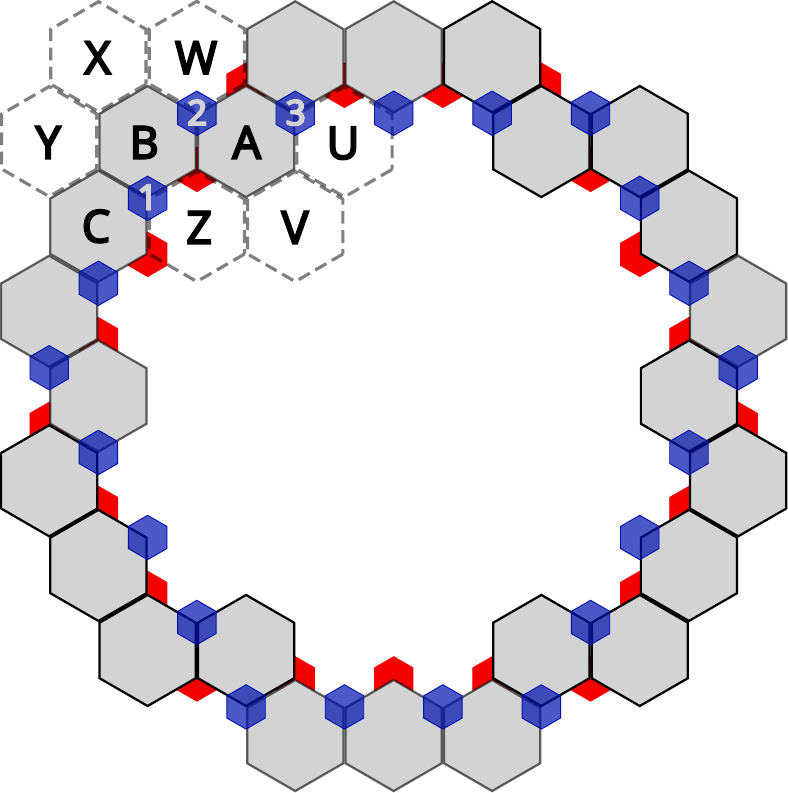}
    \caption{A super rigid configuration, positions U, V, W, X, Y, and Z are empty}
    \label{fig:SuperRigid}
\end{figure}

\begin{figure}[ht!]
    \centering
    \includegraphics[scale=.5]{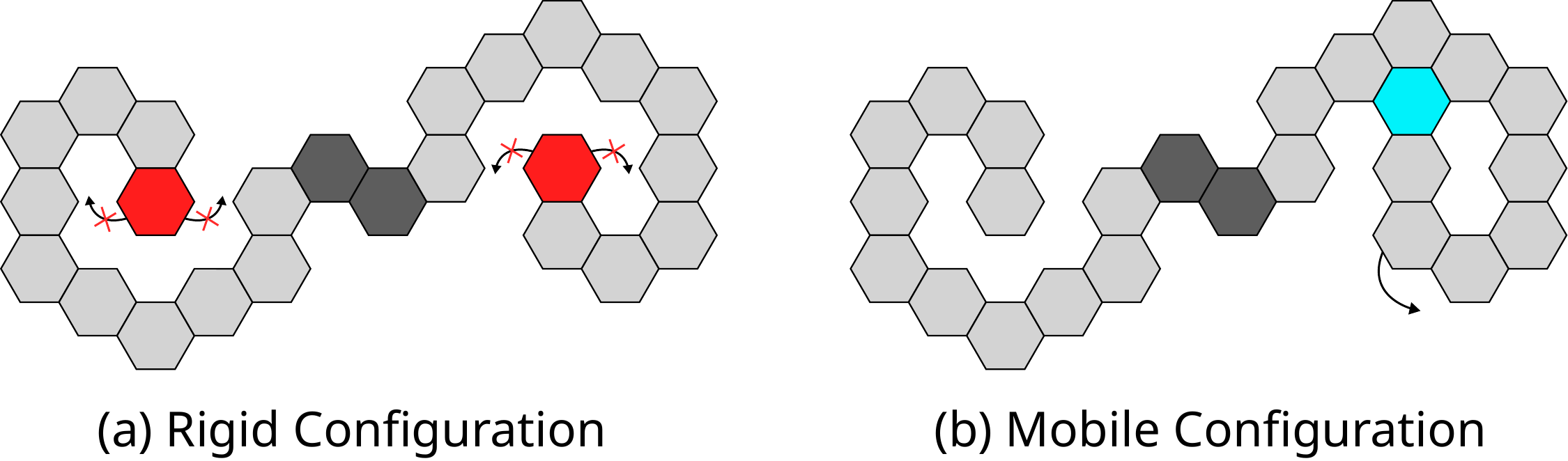}
    \caption{In (b) the addition of the blue module makes configuration (a) mobile}
    \label{fig:FreeRigid}
\end{figure}

Where a \emph{subconfiguration} of a configuration $C$ is any connected configuration that can be obtained by removing any number of modules from $C$. To our knowledge, super-rigid configurations are novel, and likely only possible in MSR models where moving modules can have collisions outside their plane of movement. Previously known rigid configurations (including all 2D rigid configurations) are like configuration (a) in Fig. \ref{fig:FreeRigid}, where adding just one new module (in blue) the configuration becomes mobile (b).

\begin{theorem} The configuration in Fig.~\ref{fig:SuperRigid} is super rigid under restricted and monkey moves.

\end{theorem}

The proof (Appendix~\ref{Super-Rigid-Appen}) uses the free-space requirements in Fig. \ref{fig:RDMoves} to verify no module in $G$ is mobile and none can be made mobile by adding any modules to $G$. A surprising consequence of this is a solid ``cube'' containing Fig. \ref{fig:SuperRigid} cannot be reconfigured into a single row of modules in a line.

\begin{corollary}
The Rhombic Dodechadral MSR model is not universally reconfigurable under Restricted or Monkey moves.
\end{corollary}




\bibliography{bibliography}

\appendix
\section{Expanded PSPACE Construction Discussion}
\label{PSPACE-Appen}


This construction has two pieces, a roof and a cap, the cube and RD variants of these are shown in Fig.~\ref{fig:roofs}. A roof configuration is a gadget contained in a single layer formed by a cycle of modules, the \emph{boundary}, and all positions that are enclosed by the boundary. A cap configuration can be placed at the end of a path of modules to make it rigid. The general idea of a cap configuration is the same for cubes and RD. A line of modules comes in from one end (shown by the dashed arrow) and then the configuration wraps around itself such that the final module cannot move and no other module can move without disconnecting the configuration. \par
In the cube cap, a path of modules makes a loop and then position its terminal module (in pink) so the top face is surrounded on all sides by other modules. Thus the module cannot move without hitting a module. The RD cap is similar, though instead of completely surrounding the terminal module, our loop puts a module one layer above and one layer below it, on directly opposite sides of the terminal module. These lock it in place, as any possible move will hit one of these two modules.\par
As a module surrounded on all sides is clearly immobile, the only mobile modules in a roof could be those on the boundary (shown in dark grey). At each boundary module we attach a path of modules running away from the roof. These need to alternate between adjacent edge modules, one path goes away horizontally, the next goes away  vertically from the roof, etc... This way no path contains a module adjacent to another path, avoiding cycles in the adjacency graph which would likely create mobile modules. After some arbitrary constant length, we end each path with a cap configuration. Now each boundary module has a collection of modules attached to it, they cannot move without disconnecting their attached modules. Every module in a path is either immobile or cannot move without disconnecting its end. Finally the internal modules in the roof configuration are surrounded and therefore immobile, so the roof configuration is immobile.   

\section{Super Rigid Case Analysis}
\label{Super-Rigid-Appen}
We show there are no modules that could be added to the configuration in Fig.~\ref{fig:SuperRigid}, call it $G$, to make it mobile. As the grey layer is symmetric there are only 3 modules we need to consider, A, B, and C. Additionally the red and blue layers are mirror images of each other so we can reduce our case analysis to the blue layer, and further we only need consider modules 1 and 2. 

\begin{itemize}
    \item[] \underline{\textbf{Modules 1 and C:}} Both of these modules have two neighbors on direct opposite sides. A module that is sandwiched like this can never move unless one of its neighbors moves first. Therefore there is no position where we could add a module to make these mobile, under either restricted or monkey moves.
    \item[] \underline{\textbf{Module B:}} There are 4 positions in the grey layer adjacent to B. If we add a module at Z or X, B gains no new options for movement as A and C block any in-layer rotation, and red and blue modules block any possible out-of-layer movement. If a module was added at either Y or W then B would have two adjacent modules on directly opposite sides, either $\{A, Y\}$ or $\{C, W\}$ making B immobile. Therefore the only remaining options are an out-of-layer move or a monkey move. B cannot make a monkey move in any direction as the existence of 1 and 2 forbid it. B cannot rotate out of the layer with a restricted move as 1 and 2 prevent it from moving up a layer or down a layer. If B tried to move down to the red layer its neighbors in the grey layer would prevent such a move. The existence of A and C also prohibit any possible face move by B.
    \item[] \underline{\textbf{Module A:}} If we add a module at Z or U then A would have two adjacent modules on directly opposite sides, making A immobile. If we added a module at position W, A would be unable to make any in-layer rotation around it as its grey neighbors would interfere. Similar to B, 2 and 3 prevent A from moving up or down a layer with a base in the grey or blue layer. If A moved down a layer using a base in the red layer, its neighbors in the grey layer would prevent such a move. Finally, A could never make a monkey move as it has two neighbors in the layers above and below it, which violates both monkey move free-space requirements.
    \item[] \underline{\textbf{Module 2:}} The modules below 2 in the grey layer prevent any restricted move 2 would make. 2 could not make a move across the face of the module below it as 1 and 3 are non-adjacent neighbors which instantly prevents this move. 1 and 3 also prevent a move up a layer with a base in the blue layer. The grey layer modules also prevent an upward or downward move using an out-of-layer base. Finally, 1 and 3 again prevent any monkey move as they are two non-adjacent neighbors of 2.
\end{itemize}

As there is no position we could add a module to make $G$ mobile, then there is no configuration with $G$ as a sub-configuration that is not at least locked if not rigid. Therefore $G$ is a super rigid configuration.
\end{document}